\documentclass[a4paper,10pt,twoside]{cpc-hepnp}
\usepackage{CJK,upgreek,fancyhdr}
\usepackage{multicol}
\usepackage{graphicx}
\usepackage{booktabs}
\usepackage{amssymb,bm,mathrsfs,bbm,amscd}
\usepackage[tbtags]{amsmath}
\usepackage{lastpage}

\begin{document}
\begin{CJK*}{GB}{gbsn}

\fancyhead[c]{\small Chinese Physics C~~~Vol. xx, No. x (201x) xxxxxx}
\fancyfoot[C]{\small 010201-\thepage}

\footnotetext[0]{Received 20 July 2016}

\title{Fast and accurate generation method of PSF-based system matrix for PET reconstruction\thanks{Project supported by National Natural Science Foundation of China (Grant No 81301348) and China Postdoctoral Science Foundation (Grant No 2015M570154)}}

\author{%
     Xiao-Li Sun(ËïУÀö)$^{1,2,3}$
\quad Shuang-Quan Liu(ÁõË«È«)$^{1,2}$
\quad Ming-Kai Yun(\hbox{\lower-1.0ex\hbox{\scalebox{1}[0.5]{ÛÌ}}\lower.1ex\hbox{\kern-1em \scalebox{1}[0.6]{±´}}}Ã÷¿­)$^{1,2}$\\
\quad Dao-Wu Li(ÀîµÀÎä)$^{1,2}$
\quad Juan Gao(¸ß¾ê)$^{1,2}$
\quad Mo-Han Li(ÀîĬº­)$^{1,2,3}$
\quad Pei Chai(²ñÅà)$^{1,2}$\\
\quad Hao-Hui Tang(ÌƺƻÔ)$^{1,2}$
\quad Zhi-Ming Zhang(ÕÂÖ¾Ã÷) $^{1,2}$
\quad Long Wei(κÁú)$^{1,2;1)}$\email{weil@ihep.ac.cn}%
}
\maketitle

\address{%
$^1$  Institute of High Energy Physics, Chinese Academy of Sciences, Beijing 100049, China\\
$^2$  Beijing Engineering Research Center of Radiographic Techniques and Equipment,
Beijing 10049, China\\
$^3$  University of Chinese Academy of Sciences, Beijing 100049, China\\
}

\begin{abstract}
 Positional single photon incidence response (P-SPIR) theory is researched in this paper to generate more accurate PSF-contained system matrix simply and quickly. The method has been proved highly effective to improve the spatial resolution by applying to the Eplus-260 primate PET designed by the Institute of High Energy Physics of the Chinese Academy of Sciences(IHEP). Simultaneously, to meet the clinical needs, GPU acceleration is put to use. Basically, P-SPIR theory takes both incidence angle and incidence position by crystal subdivision instead of only incidence angle into consideration based on Geant4 Application for Emission Tomography (GATE). The simulation conforms to the actual response distribution and can be completed rapidly within less than 1s. Furthermore,two-block penetration and normalization of the response probability are raised to fit the reality. With PSF obtained, the homogenization model is analyzed to calculate the spread distribution of bins within a few minutes for system matrix generation. As a reult, The images reconstructed indicate that the P-SPIR method can effectively inhibit the depth of interaction (DOI) effect especially in the field close to the edge of the field of view (FOV). What is more, the method can be promoted to any other PET and the list-mode organization structure high-speedily and efficiently, which substantially reduces the computing cost and ensures the accuracy of system matrix for PET reconstruction.
\end{abstract}



\footnotetext[0]{\hspace*{-3mm}\raisebox{0.3ex}{$\scriptstyle\copyright$}2016
Chinese Physical Society and the Institute of High Energy Physics
of the Chinese Academy of Sciences and the Institute
of Modern Physics of the Chinese Academy of Sciences and IOP Publishing Ltd}%

\begin{multicols}{2}

\section{Introduction}
Positron emission tomography increasingly plays a vital role in the modern nuclear medicine with the technical progress and maturity of both the detectors and reconstruction algorithms.\cite{lab1}As a kind of functional imaging methods, PET is widely used in basic medical research, clinical diagnosis, new drug research and development, the curative effect evaluation and so on. PET uses the positron-electron annihilation, which can be considered producing a pair of photons with same energy but opposite direction. Each photon will be detected by the crystal array as a single, and the two singles can compose to a line of response (LOR) called coincidences to form data reconstructed.However, the depth-of-interaction (DOI) can result in positioning error which reduces the spatial resolution and image accuracy.\cite{lab2,lab3} Multilayer-crystal detector structure has been adopted to decrease the DOI effect widely.But, it increases cost of materials and processing. Point spread function (PSF) describing response distribution of a point source is considered to be effective method to weaken the DOI effect in the reconstruction, which makes the spatial resolution improved.\cite{lab4}Traditionally, the PSF can be acquired from experimental measurement,\cite{lab5}Monte Carlo simulation\cite{lab6}or analytical derivations\cite{lab7}. However, the precise of orientation for point source in the experimental measurement, the enormous cost of computing time and computing resource in the simulation or the accuracy and rationality of the analytical methods become the bottlenecks of traditional methods. Instead of constrains, a single photon incidence response theory(SPIR) has been proposed with our team in (Fan Xin 2015)\cite{lab8} creatively with focus on single photon behavior. Nevertheless, it still has insufficiency in accuracy of the model and the process of gap between two blocks. Further work is continued all the time.

In this work, positional single photon incidence response (P-SPIR) developed out of SPIR is studied to obtain point spread function. The penetration effect happening in two neighbor blocks is fully considered. And a homogenization model is provided to generate system matrix based on the response of single photon. In succession, system matrix based on PSF can be generated for PET reconstruction. It has been proved to improve the image spatial resolution with less computing cost and higher generating speed.

\section{Methods}

Based on the common ground of the single photon incidence, this paper puts forward a comprehensive method with both simulation and analytical calculation to make system matrix accurate and generated high-speedily for PET reconstruction.

\subsection{Positional SPIR Model}

The SPIR theory takes a starting point that the simulation result can be equally applicable in other PET system when the crystal size is the same. It makes the simulation work separated from the concrete structure of the detector. However, in fact, the SPIR model just considering incidence angle cannot describe the response distribution precisely. Constantly, the positional single photon incidence response (P-SPIR) model takes both incidence angle and incidence position into account in the simulation. Furthermore, the P-SPIR theory solves the problem of the gap influence when the penetration effect happens in two neighboring blocks based on analytical calculation. It has been proved to significantly and foreseeably reduce the computing cost under the circumstance of the system matrix well generated and image quality improved.

\subsubsection{Response Model}

It comes to light that the penetration effect determines the response distribution other than the incidence position where the gamma photon injects into the crystal array. In the previous paper, the point to discuss the response distribution after the incidence always lies in the angle of incidence. As in SPIR model, the angle of incidence is simulated from 0 to 60 degree in 5-degree interval as Fig.1. However, there will be significant differences taking the incidence position into account as Fig.2. The P-SPIR response model divides each discrete crystal bar into x equal parts. In Fig.2, the 7th crystal bar is allocated into eight equal segments(\emph{x}=8). The details are studied with same angle of incidence but different incidence position in one crystal bar as Fig.3 on the simulation platform of GATE\cite{lab9} by controlling the incidence direction of the point source with biased source. The number of source particles is \emph{M}, the count number of the crystal bars is $\emph{M}_{\emph{i}}$ (\emph{i}=0, 1, 2...15). So the response probability of the crystal \emph{i} is
\begin{eqnarray}
\label{eq1}
{\rho _i} = \frac{{{M_i}}}{M} \times 100\% \
\end{eqnarray}

When the $\alpha$ in Fig. 2 equals to 30 degree, the curve AA* and BB* can show the response distribution of point A and B in Fig. 3. On the overall view of the data in Fig. 3, the peak position and peak value differs with the change of incidence position in one crystal bar. On one side, the peak position changes. For the curve p0, the maximum lies on the 7th crystal while others achieve maximum on 8th crystal. On the other side, the p1, p2¡­p7 share with the same peak position but definitely different distribution. The accuracy of the distribution is proved of much importance in the calculation of the system matrix.
\begin{center}
\includegraphics[width=5cm]{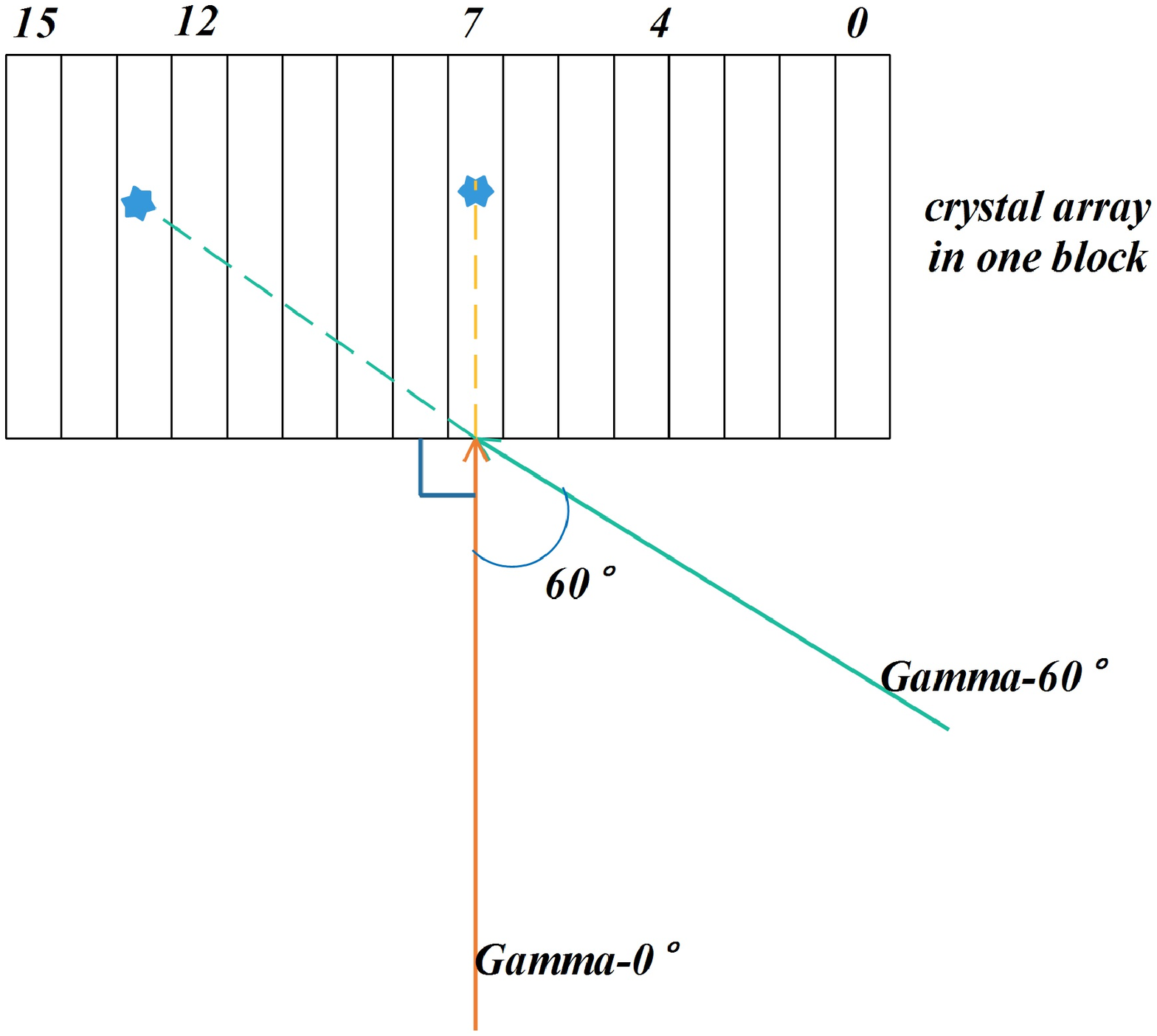}
\figcaption{\label{fig1}   The response model of SPIR. }
\end{center}
\begin{center}
\includegraphics[width=6cm]{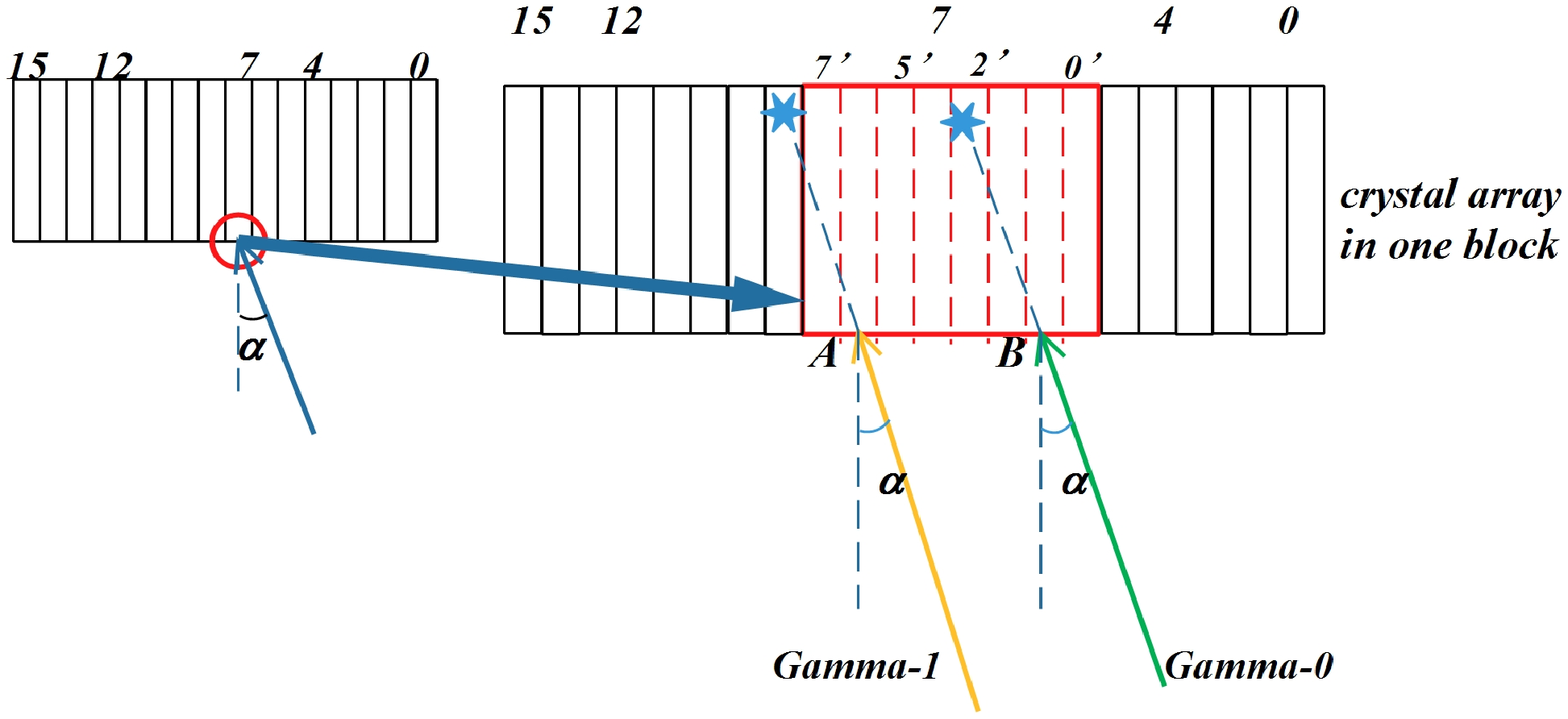}
\figcaption{\label{fig2}   The response model of P-SPIR. }
\end{center}
\begin{center}
\includegraphics[width=5cm]{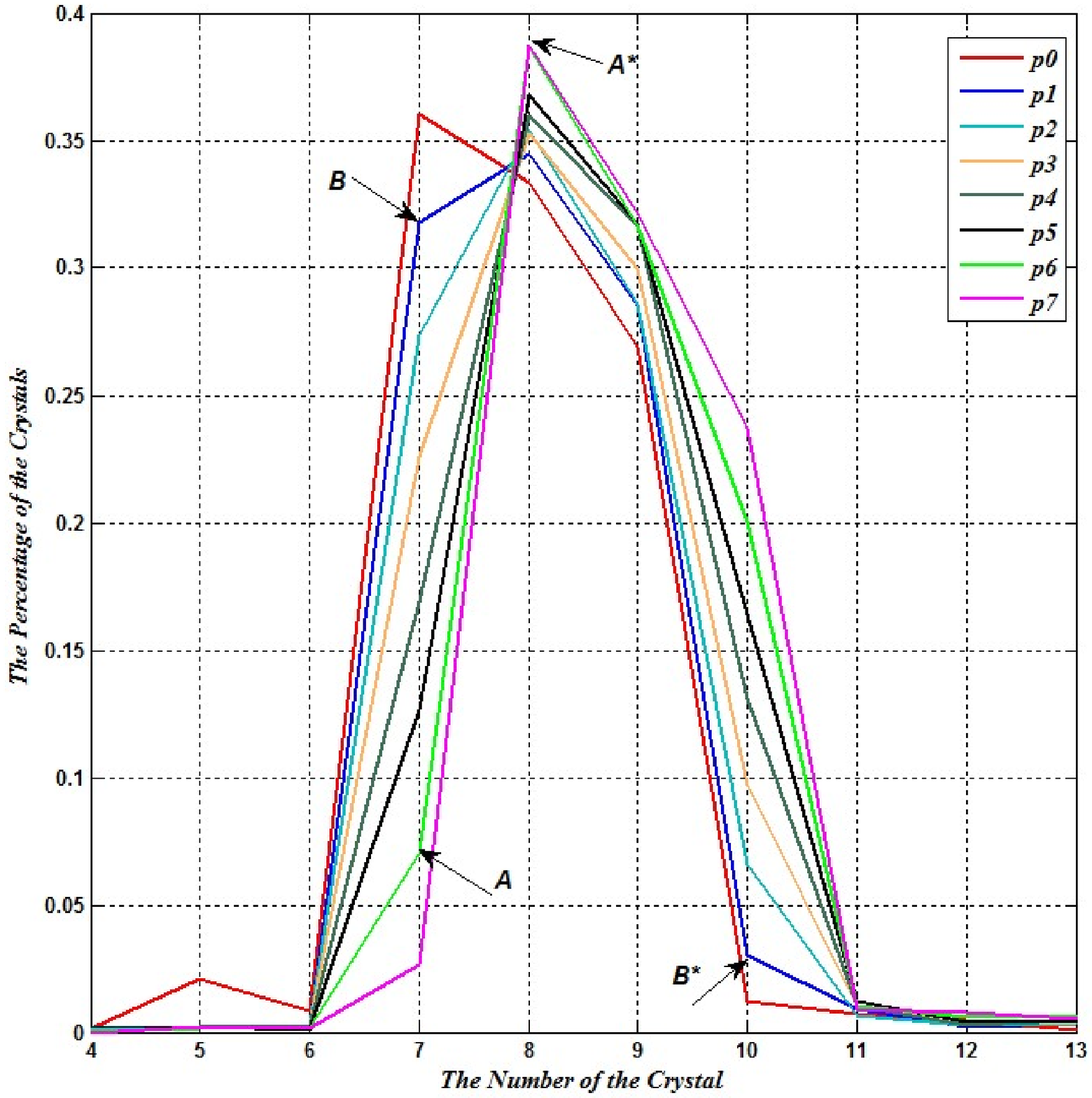}
\figcaption{\label{fig3}    The response distribution of different incidence position with the same incidence angle. }
\end{center}
\subsubsection{The Process of the Gap Influence}
On the edge of the block, the penetration effect may happen in two neighboring blocks in consideration of the gamma photon going through from one block to another as Fig. 4 shows. In the SPIR model, the particularity with the gap is almost ignored. After further study by GATE simulation method, the gamma photon arriving at the next block can be in a large probability at relatively large incidence angles as Fig. 5. It is assumed that the order number of one block is from 0 to 15 and 16 to 31 is for the next block. Thus, the 15th and the 16th are close to each other divided by the gap. The curve means different incidence angles at the same incidence position of the 7th position of the 15th crystal. From the curves changing trends, it is clear that the penetration effect leads to the dominant position of the neighboring block with crystals from 16 to 31 other than the injected block with crystals from 0 to 15. As a result, the penetration effect between neighboring blocks should be considered carefully.
\begin{center}
\includegraphics[width=5cm]{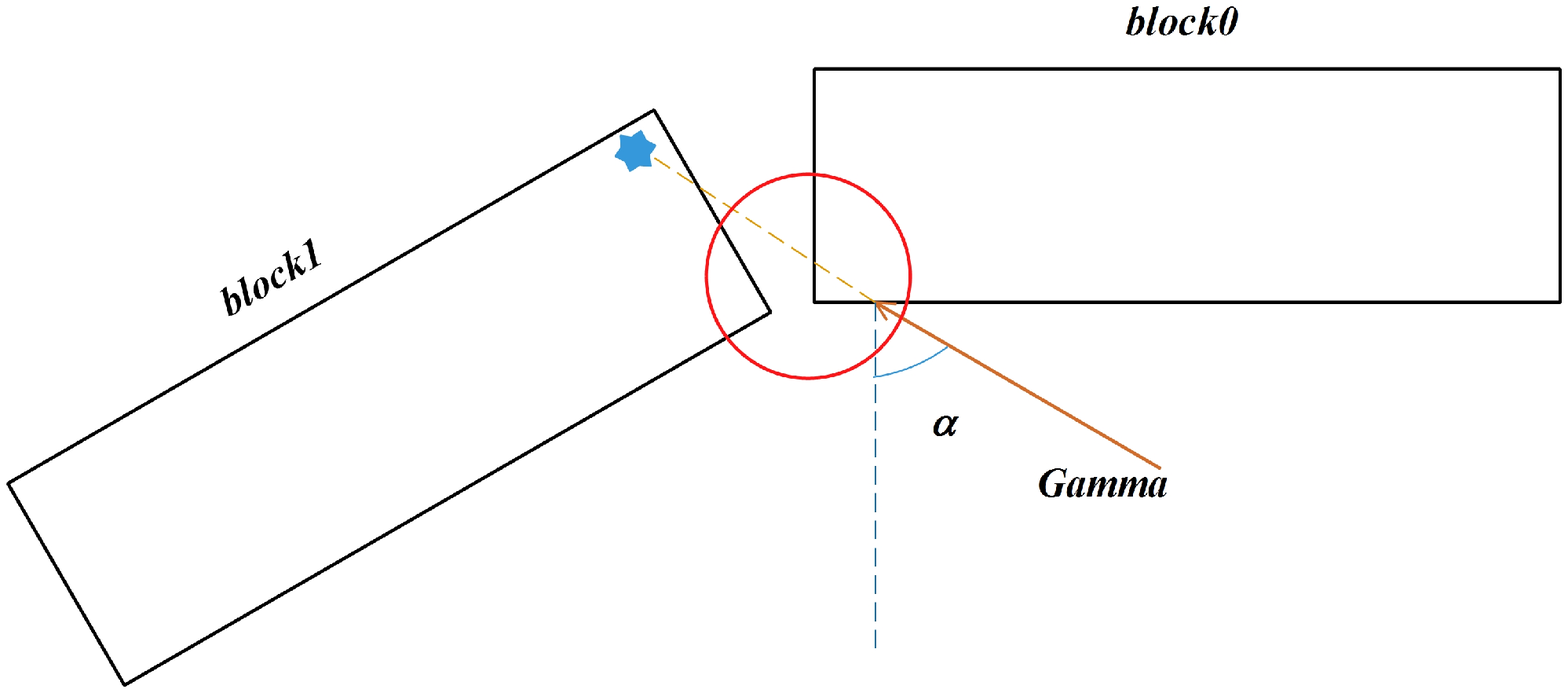}
\figcaption{\label{fig4}  The gap between two blocks. }
\end{center}
\begin{center}
\includegraphics[width=5cm]{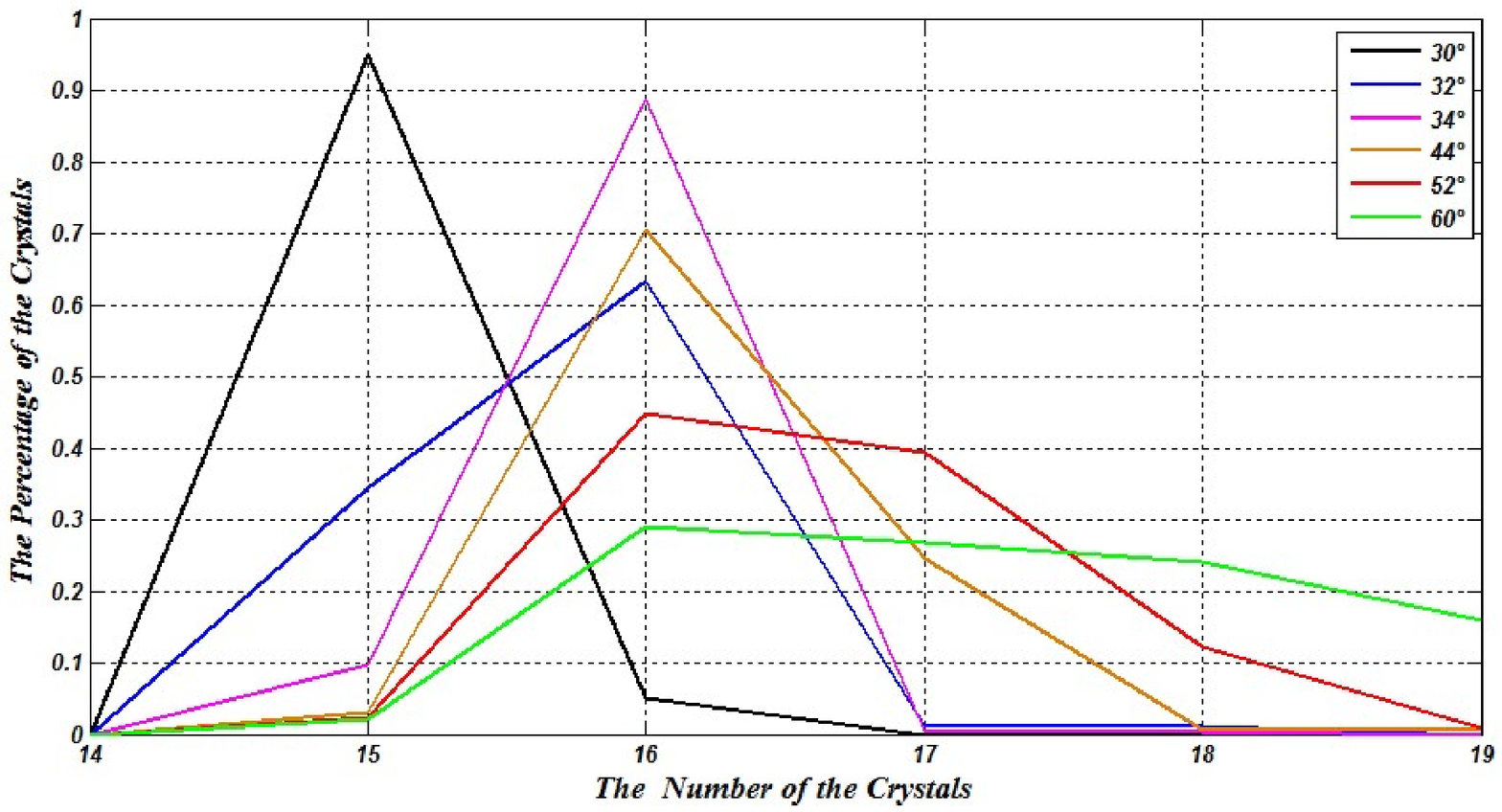}
\figcaption{\label{fig5} The distribution of two neighboring block. }
\end{center}

When the gap between two neighboring blocks taken into account, the P-SPIR theory will become invalid in separating from the concrete detector structure, which fades the superiority and the novelty in the simulation. So the gap relationship marked by a circle in Fig. 4 is solved with a method of analytical calculation is provided as Fig. 5 shows.

The formula of the equivalent incidence angle into the next block $\beta$ is
\begin{eqnarray}
\label{eq2}
\beta  = \alpha  - \frac{{2\pi }}{R}\
\end{eqnarray}
Where \emph{R} is the block number in one ring, $\alpha$ is the original incidence angle.

The formula of the equivalent incidence position into the next block $\emph{AP}^1$ is
\begin{eqnarray}
\label{eq4}
A{P^1}{\rm{ = }}\sigma {\rm{ + }}\frac{{sin({\pi  \mathord{\left/
 {\vphantom {\pi  2}} \right.
 \kern-\nulldelimiterspace} 2} - \alpha )}}{{sin({\pi  \mathord{\left/
 {\vphantom {\pi  2}} \right.
 \kern-\nulldelimiterspace} 2} + \beta )}}(\sigma  + \nu )\
\end{eqnarray}
\begin{eqnarray}
\label{eq5}
n = \frac{{A{P^1}}}{\lambda }\
\end{eqnarray}
Where $\lambda$ is the width of the crystal transaxially,$\sigma$ is the length of the gap size like \emph{AB} and \emph{BC} in Fig. 6,$\nu$ is the distance from the incidence position to the edge as $\emph{CP}^0$. The decimal path of \emph{n} is the proportion of crystal bar injected for the incidence position.

When the incidence angle $\beta$ and the incidence position \emph{n} are obtained, the distribution response of the block1 can be calculated as the response model described in 2.1.1.
\begin{center}
\includegraphics[width=6cm]{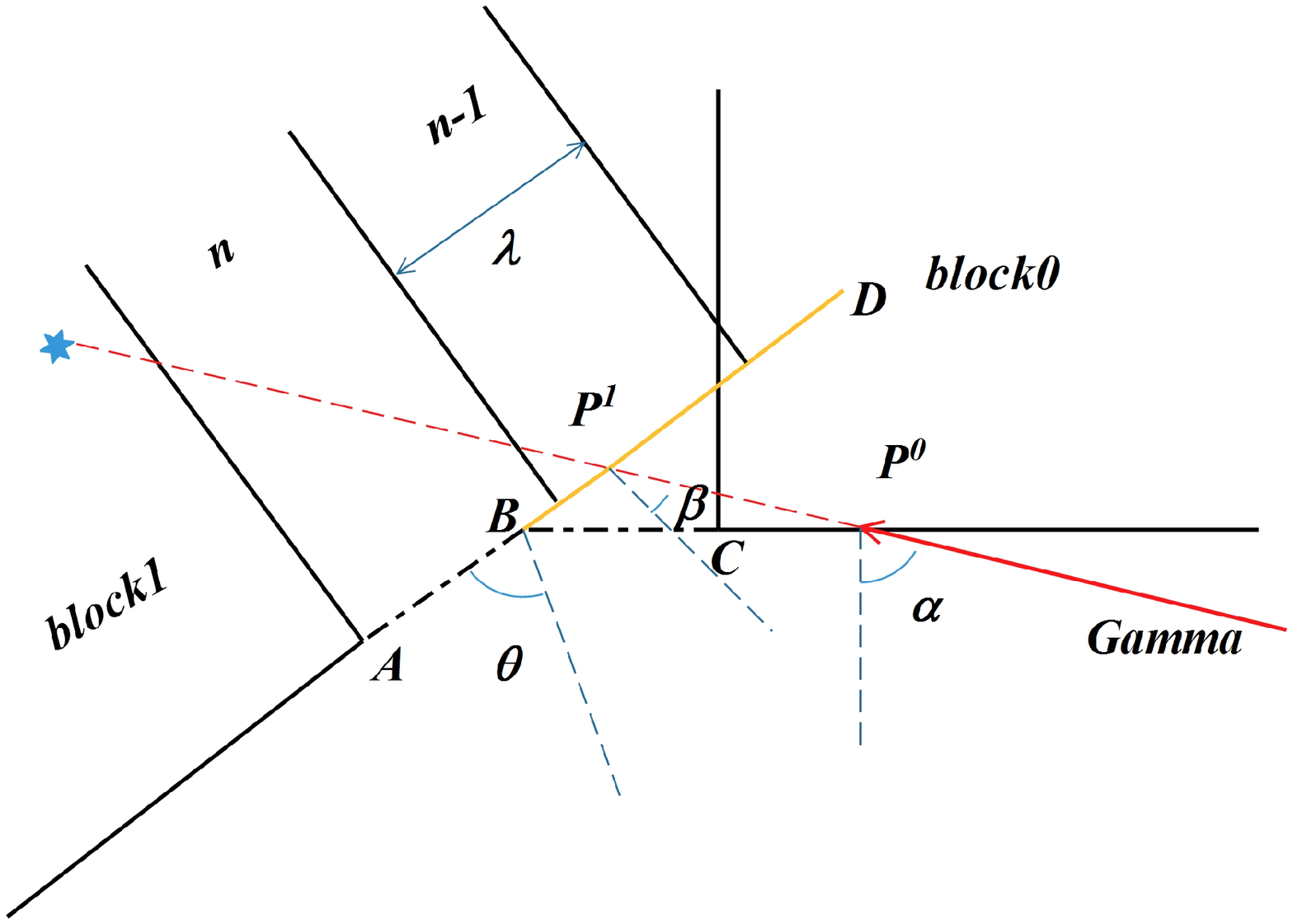}
\figcaption{\label{fig6} The approximate calculation model of the gap. The \emph{AB} and \emph{BC} means the gap between two neighboring blocks. The \emph{BD} is the extension line of the \emph{AB}. The \emph{n}th crystal bar is suppositional part generated for the block1. }
\end{center}

\subsubsection{Normalization}
In the simulation, the source activity is fixed with a same simulation time in spite of different incidence directions and positions. However, affected by the specific physical factors, the counting rate comes to significantly different. Thus the response distribution values should combine with the detection efficiency respectively. For example, the source activity is \emph{A}. the simulation time is \emph{t}. The number of effective singles detected totally is \emph{M}. So the formula of the counting rate $\eta$ is
\begin{eqnarray}
\label{eq6}
\eta {\rm{ = }}\frac{M}{{At}} \times 100{\rm{\% }}\
\end{eqnarray}
The response probability of one crystal bar at a kind of incidence angle and position in the whole PET system comes to be the formula as follows.
\begin{eqnarray}
\label{eq7}
\rho {\rm{ = }}\eta {\rho _0}\
\end{eqnarray}
Where $\rho_0$ is the response probability calculated in Eq.(1).

\subsection{Probability Calculation}
In the PET system, the point lies on the LOR's distribution. A LOR consists of a pair of two gamma photons with the same energy but different direction generated at the same time by annihilation reaction. The LOR's proportion in the whole system is
 \begin{eqnarray}
\label{eq10}
{\rho _{LOR}} = {\rho _1}{\rho _2}\
\end{eqnarray}
Where $\rho_1$, $\rho_2$ are the proportions of the two gamma photons.

\subsection{System Matrix Generation}
The P-SPIR model describes the response distribution when a gamma photon injecting into a crystal array by both simulation and calculation. Here, a homogenization model is provided to generate the system matrix voxel by voxel.

The system matrix comes from the LOR distribution of the voxels based on the dimension of the reconstruction image. The homogenization theory makes a hypothesis that an ideal point source emits isotropic gammas with equal probability in the whole space. The ideal point source is with its geometric volume ignorable and placed in the center of the voxel as Fig. 7. The proportion of $LOR_1$, $LOR_2$, $LOR_3$ and $LOR_4$ are the same. The angle between the $LOR_\emph{i}$ and the horizontal axis is
 \begin{eqnarray}
\label{eq11}
{\theta _i} = \frac{\pi }{k}i,\left( {i = 0,1,2...k - 1} \right)\
\end{eqnarray}
Where the \emph{k}is the number of the LORs in the model. When the \emph{k} is large enough, the LORs can be deemed to be homogeneous. With the increase of \emph{k}, the computing cost becomes larger. But the result approaches the steady state at a certain value.
\begin{center}
\includegraphics[width=5cm]{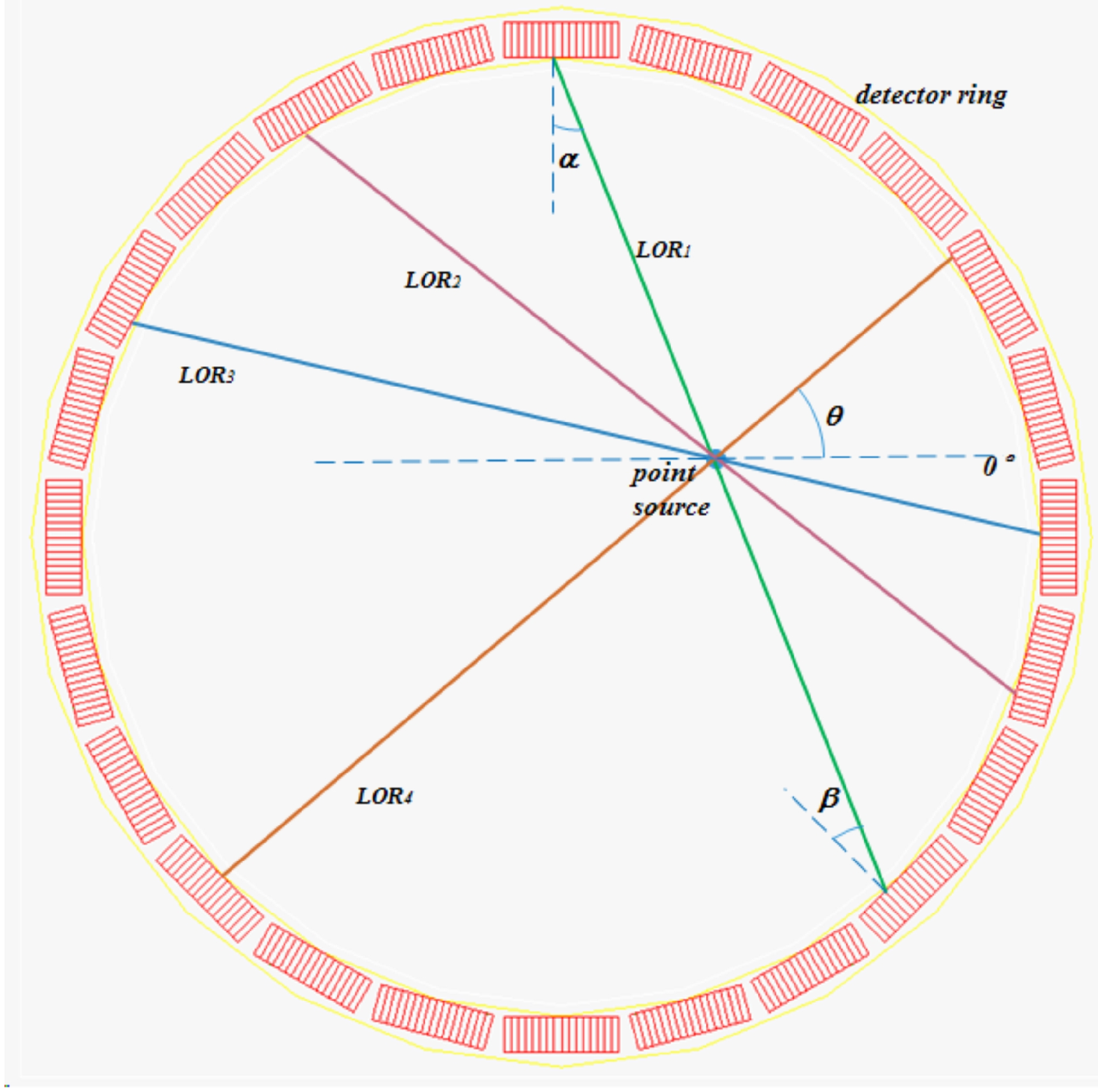}
\figcaption{\label{fig7}   The homogenization model. }
\end{center}

Specifically speaking, the homogeneous emission LOR above will be detected as a few of LORs account for the penetration effect as Fig. 8 shows. For the LOR generated by the homogeneous model, the incidence crystal number is \emph{f} in block0 and \emph{n}¡¯ in block1. As a result of point spread function, the crystals \emph{b} to \emph{f} in block0 and \emph{o}¡¯ to \emph{f}¡¯ can put out detection signals. So the detected LORs can be the combination of \emph{b} to \emph{f} and \emph{o}¡¯ to \emph{f}¡¯. The LOR¡¯s proportion in the whole system can be calculated as Eq.(12). Specially, when the gamma goes through the gap, the proportion is zero.

Ultimately, the proportion of a LOR is
 \begin{eqnarray}
\label{eq12}
{\rho _{LOR}} = \sum\limits_0^{N - 1} {{\rho _i}} {\rho _j}\
\end{eqnarray}
Where the \emph{i}th and \emph{j}th crystal combines the LOR.
\begin{center}
\includegraphics[width=7cm]{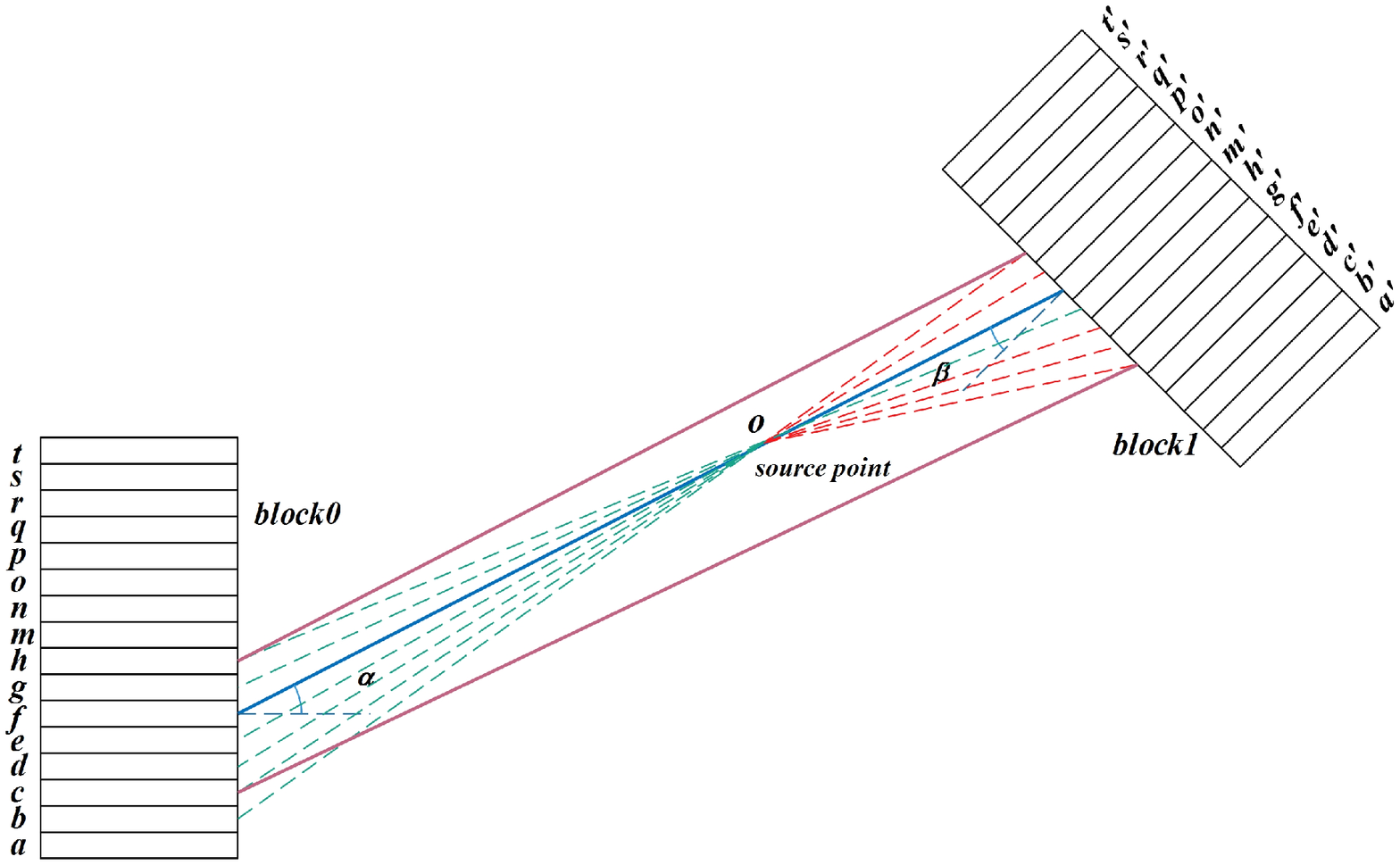}
\figcaption{\label{fig8}   The spread of LORs. }
\end{center}

\subsection{The Comparison of P-SPIR and Simulation Method }

The simulation voxel by voxel is a traditional method to generate system matrix.\cite{lab10,lab11}The voxel-by-voxel simulation takes an enormous computing time to meet the data needs and completely depends on the PET structure. It has been proved that the computing cost can be huge even though the symmetry is taken into a full consideration.\cite{lab12,lab13}Meanwhile, it is assessed that the computing time of P-SPIR can be much less than 1s with a single core computer. Furthermore, it is much more meaningful than the time saving that the system matrix can directly generate with the detector settings based on the response distribution of the P-SPIR already simulated with the PET settings if the crystal size is the same. Consequently, the P-SPIR has significant advantages in system matrix generation.

\section{Results}
The research is carried out on the Eplus-260 Primate PET designed by the Institute of High Energy Physics of the Chinese Academy of Sciences to be suitable for the functional imaging of the head of primates and the whole body of rodents. The scanner is provided with good spatial resolution and detection efficiency containing of high-performance LYSO crystals and position sensitive photomultiplier tubes. The scanner comes to two rings with 24 detector modules each. A module contains of an array of 16¡Á16 crystals with dimensions of 2mm¡Á2mm¡Á10mm. The system has 768 crystals in total and 32 rings with a pitch of 2mm. The field of view (FOV) is a cylinder of 190mm in diameter transaxially and 64mm in height axially. Based on the Eplus-260 Primate PET scanner, the image space and the LOR histogram configuration are listed in Table 1.

Both simulation data and experimental data are reconstructed with OSEM. The full width at half maximum(FWHM) of simulated points and reconstruction image of experimental Derenzo phantom are analyzed.
\begin{center}
\tabcaption{ \label{tab1}  Image Volume and LOR histogram parameters.}
\footnotesize
\begin{tabular*}{65mm}{c@{\extracolsep{\fill}}ccc}
\toprule Settings &Value \\
\hline
Number of angles\hphantom{00} & \hphantom{0}192 \\
bins per angle\hphantom{00} & \hphantom{0}199 \\
Image voxels\hphantom{00} & \hphantom{0}380¡Á380¡Á63 \\
Voxel size ($mm^3$)\hphantom{00} & \hphantom{0}0.5¡Á0.5¡Á1.0 \\
\bottomrule
\end{tabular*}
\end{center}

\subsection{FWHM of Point Sources }
Point source at the radial position of (5, 10, 15, 25, 50 and 75) mm and zero-offset in the tangential direction off center as National Electrical Manufactures Association(NEMA) is simulated by GATE. The FWHW of the reconstruction image is analyzed as Fig. 9. On one hand, the reconstruction image of the point source close to the edge becomes significantly expanded without PSF. On the other hand, the FWHM of the point with PSF is much smaller especially at the position of 50mm and 75mm off center, which means the DOI effect is obviously weakened in edge regions.
\begin{center}
\includegraphics[width=7cm]{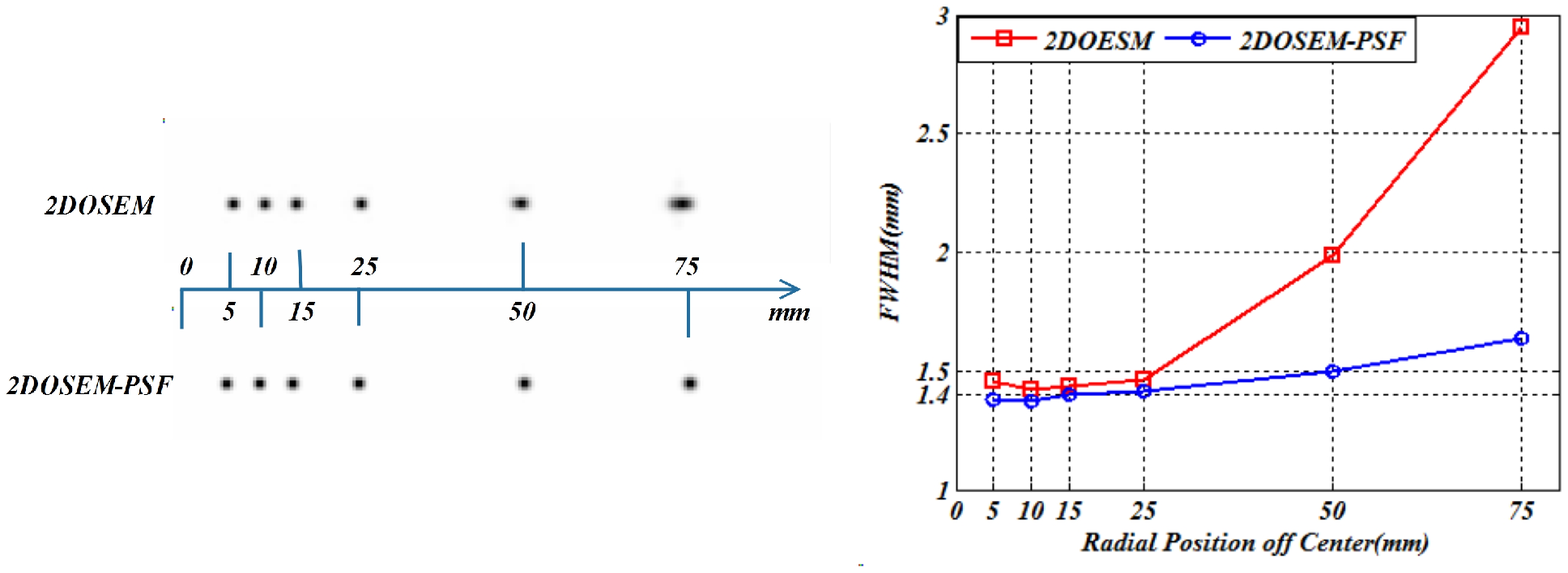}
\figcaption{\label{fig9}   The reconstruction image of point sources. }
\end{center}

\subsection{Derenzo Phantom }
The Derenzo phantom as Fig. 10 is used to verify the spatial resolution of the Eplus-260 Primate PET system. The image reconstruction of OSEM, OSEM-PSF with Monte Carlo and OSEM-PSF with P-SPIR are displayed as Fig. 11. It shows the 40th slice reconstructed with 20 times of iteration and 24 subsets, which clearly indicates that the reconstruction method with PSF prevails in distinguishing the point sources in the 1.35mm-field compared to the none-PSF method. Furthermore, from the profiles, the curve troughs with PSF are much more well-defined than the 2DOSEM without PSF. Meanwhile, the result derived from method with P-SPIR is consistent with the method with Monte Carlo simulation.
\begin{center}
\includegraphics[width=5cm]{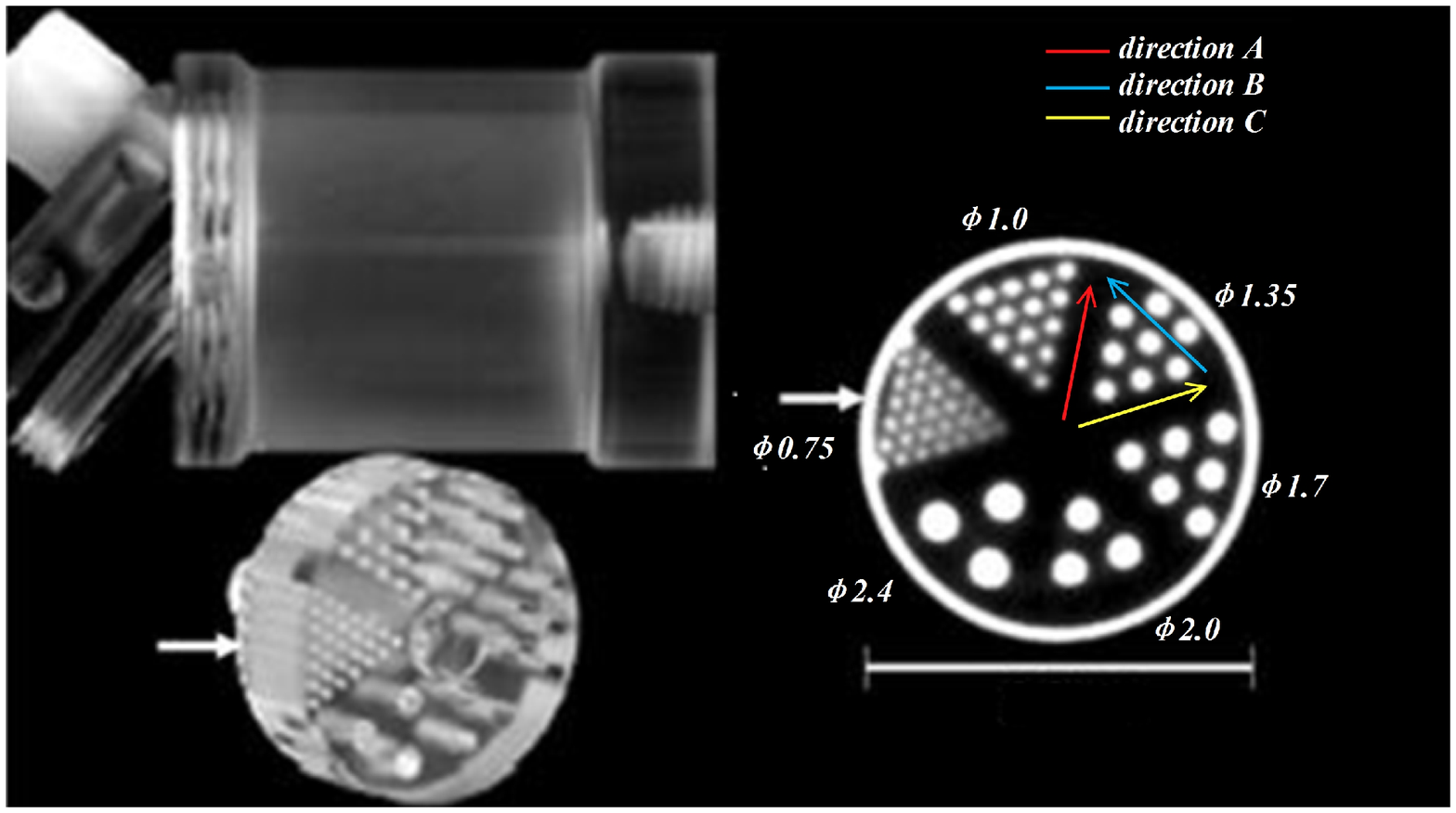}
\figcaption{\label{fig10}   Derenzo phantom used in the experiment. }
\end{center}
\begin{center}
\includegraphics[width=8cm]{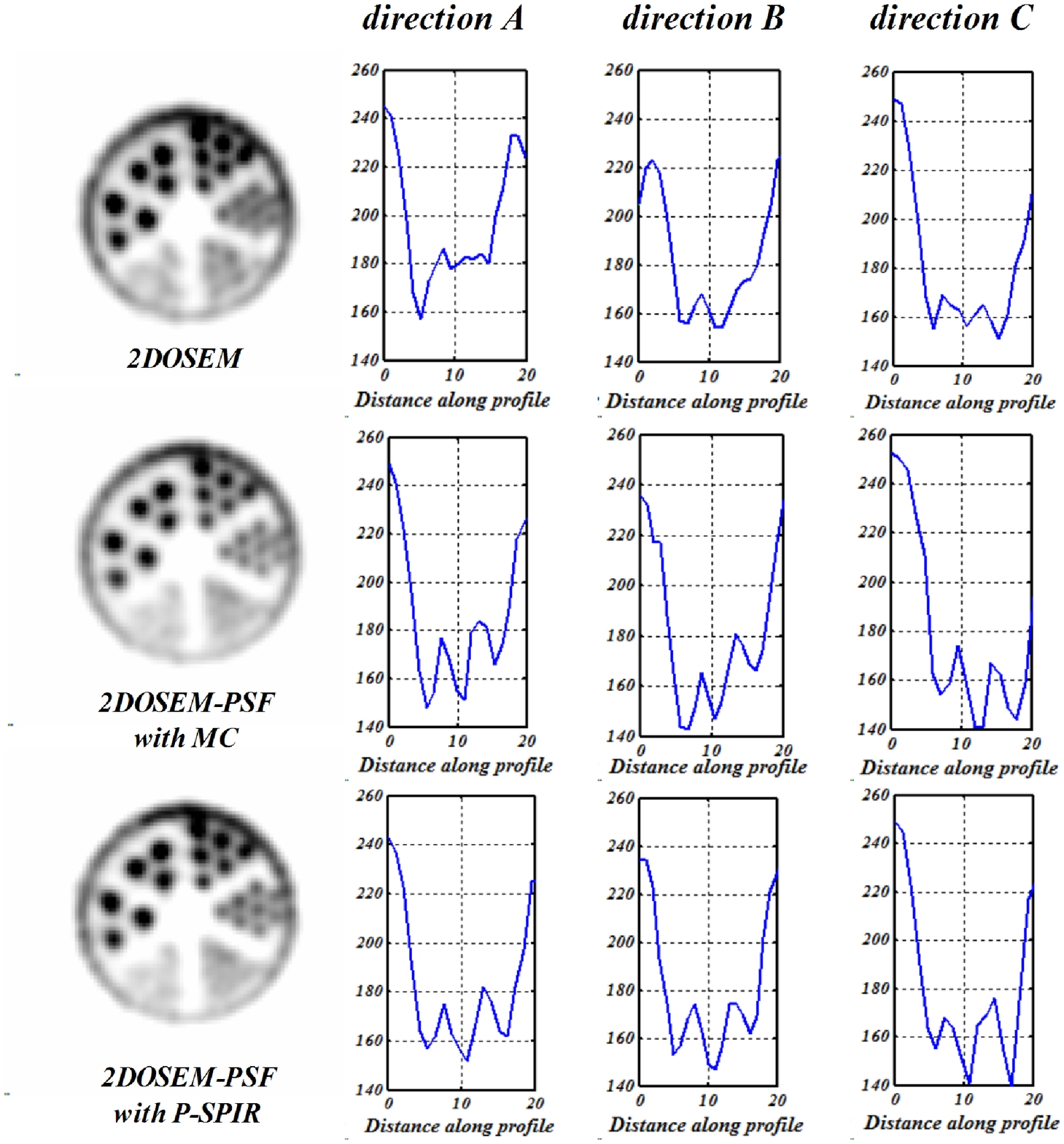}
\figcaption{\label{fig11}   The reconstruction image of the Derenzo phantom. }
\end{center}

Nevertheless, the reconstruction with PSF can cost a long running-time. The parallel computing power of GPU is utilized to reduce the reconstruction time.The comparison of the methods with CPU and GPU is presented in Table 2. The computing time of the GPU is just less than 10 percent of the CPU, which fits the clinical needs.
\begin{center}
\tabcaption{ \label{tab2}  The comparison of the PSF methods with CPU and GPU for (380¡Á380¡Á63) with once iteration under equal conditions.}
\footnotesize
\begin{tabular*}{70mm}{c@{\extracolsep{\fill}}ccc}
\toprule reconstruction method &CPU&GPU \\
\hline
computing time\hphantom{00} & \hphantom{0}479 s & \hphantom{0}43 s\\
\bottomrule
\end{tabular*}
\vspace{0mm}
\end{center}
\vspace{0mm}

\subsection{The Truncation of the Histogram Bins }
At a certain voxel, there will be 199 bins for each angle. Mostly, the proportion is close to zero. Foreseeably, the use of all the bins in the PET reconstruction can result in huge computing cost which will be extremely difficult to fulfill the requirements of the clinical application. Accordingly, the number of the extended bins of each angle at a certain voxel should be truncated in a descending order with \emph{N}. The number \emph{N} has a balance point between the accuracy of the system matrix and the computing cost. The reconstruction image compared with different value of \emph{N} shows as Fig.12 and the reconstruction time cost with GPU is provided in Table 3.
\begin{center}
\includegraphics[width=8cm]{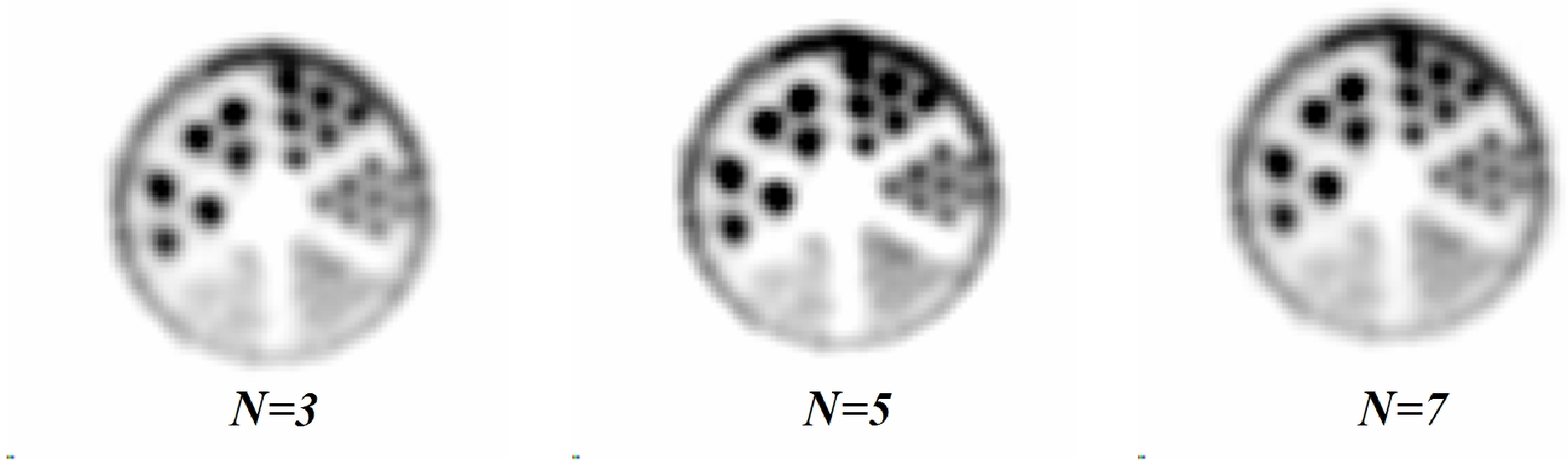}
\figcaption{\label{fig12}The reconstruction comparison with different \emph{N} at 40th slice.}
\end{center}

The Fig.12 indicates that the results are almost the same with 3, 5, 7.However, the reconstruction time is much longer with 7. In application on Eplus-260 Primate PET, 5 is chosen as a conservative approach.
\begin{center}
\tabcaption{ \label{tab3}The reconstruction time of image comparison with different \emph{N}.}
\footnotesize
\begin{tabular*}{70mm}{c@{\extracolsep{\fill}}ccc}
\toprule N &reconstruction time with GPU \\
\hline
3\hphantom{0} & \hphantom{0}38 s\\
5\hphantom{0} & \hphantom{0}43 s\\
7\hphantom{0} & \hphantom{0}113 s\\
\bottomrule
\end{tabular*}
\vspace{0mm}
\end{center}
\vspace{0mm}

\section{Discussion and Conclusions}
The system matrix comes first in the iterative methods of PET reconstruction. But the generation of the system matrix can be a huge project with experimental or simulation method accompanied of enormous computing cost and massive storage capacity. In this paper, a positional SPIR method named P-SPIR is raised with careful process of the gap and a homogenization model to improve the accuracy of the system matrix and reduce the computing cost both in the computing time and the storage capacity. What¡¯s more, the analytical calculation of the gap process and the system matrix generation based on homogenization model reinforce the reusability, portability and interoperability in various PET systems especially when the crystal size is the same. To be specific, a database can be derived from the S-SPIR mode with wide-ranging size types and all sorts of crystal materials. As a result, the system matrix for a particular PET system can be generated rapidly and efficiently on-line just once with system settings. The advantage in spatial resolution improvement has been presented in both simulation data and experimental data of the Eplus-260 Primate PET. However, several problems should be discussed in application. Firstly, though the method has been applied on Eplus-260 Primate PET and the results indicate improved spatial resolution, more support with other PET system may be needed to strength the extensive applicability and great convenience in usage. Secondly, with mentioned advantages, inferiority is potential to exist in intensity, uniformity and other performance of the reconstructing image. Thirdly, the influence of the selection of subdivision number for the crystal is better to be researched in further and in details. Lastly, the correction should be taken into full consideration to acquire images more conducive to diagnosis and treatment.

In addition, longitudinal study will be developed in two main aspects. The reconstruction results shown in this paper are based on the data organization structure of histogram. However, the system matrix generation method can also be applied in the list-mode reconstruction with a different organization structure. The list-mode reconstruction with P-SPIR will be the next step in the research. For another, in further research, a three-dimension PSF-contained system matrix will be analyzed.

\end{multicols}

\vspace{-1mm}
\centerline{\rule{80mm}{0.1pt}}
\vspace{2mm}

\begin{multicols}{2}

\end{multicols}

\clearpage
\end{CJK*}

\begin{thebibliography}{90}

\vspace{3mm}

\bibitem{lab1}Shan Tong A. M, Alessio K, Thielemans C, Stearns S, Ross P. E, Kinahan P. E. IEEE Transactions on Nuclear Science, Vol.58(5), p.2264-2275(2011)

\bibitem{lab2}Salvador S, Huss D, Brasse D. IEEE Transactions on Nuclear Science, Vol.56(1),p.17-23(2009)

\bibitem{lab3}Ito M, Lee J.S, Kwon S.I. IEEE Transactions on Nuclear Science, Vol.57(3),p.976-981(2010)

\bibitem{lab4} K. Lee, P. E. Kinahan, J. A. Fessler, R. Miyaoka, M. Janes, and T. K. Lewellen. Phys. Med. Biol. Vol.49, p.4563-4578(2004)

\bibitem{lab5}A. Alessio,C. Stearns, S. Tong, S.Ross, S.Kohlmyer, A. Ganin, and P.Kinahan. IEEE Trans. Med. Imaging, Vol.29(3), p.938-949 (2010)

\bibitem{lab6} A. M. Alessio, P. E. Kinahan, and T. K. Lewellen. IEEE Trans. Med. Imaging, Vol. 25(7), p.828-837(2006)

\bibitem{lab7} D. Strul, R. B. Slates, M. Dahlbom, S. R. Cherry, and P. K. Marsden. Phys. Med. Biol., Vol. 48(8), p. 979-994(2003)

\bibitem{lab8} Fan X, Wang H P, Yun M K, Sun X L, Cao X X, Liu S Q, Chai P, Li D W, Liu B D, Wang L, Wei L. Chinese Physics B.2015.1 542-500(2015)

\bibitem{lab9} http://www.opengatecollaboration.org

\bibitem{lab10} Vilardi I, Ciocia F, Colonna N, De Leo R, gamba L, arrone S, Nappi E, Tagliente G, Valentini A, Braem A, Chesi E, Joram C, Seguinot J, Weilhammer P, Corsi F, Dragone A, Cusanno F, Garibaldi F, Zaidi H. Nuclear Science Symposium/Medical Imaging Conference 2004.

\bibitem{lab11} Leroux J.D, Thibaudeau C, Lecomte R, Fontaine R.2007 IEEE Nuclear Science Symposium Conference Record, Vol.5, p.3644-3648(2007)

\bibitem{lab12} Long Zhang, Steven Staelens, Roel Van Holen, Jan De Beenhouwer, Jeroen Verhaeghe, Iwan Kawrakow, and Stefaan Vandenberghe. Medical Physics 37, 3667 (2010)

\bibitem{lab13} Zhang Long, Vandenberghe Stefaan, Staelens Steven, Verhaeghe J, Kawrakow Iwan, Lemahieu Ignace. 2008 IEEE Nuclear Science Symposium Conference Record,p.5101-5106(2008)

\end{thebibliography}
\end{document}